\documentclass[aps,prb,twocolumn,superscriptaddress,longbibliography]{revtex4-2}
\usepackage{epsfig,amsopn}
\usepackage{graphicx}
\usepackage{color}
\usepackage{physics}
\usepackage{amsmath,amssymb}
\usepackage{enumerate}
\newcommand\bea{\begin{eqnarray}}
\newcommand\eea{\end{eqnarray}}
\newcommand\beq{\begin{equation}}
\newcommand\eeq{\end{equation}}

\def\nn{\nonumber}

\def\al{\alpha}

\def\ep{\epsilon}

\def\ga{\gamma}

\def\si{\sigma}

\def\De{\Delta}
\def\dg{\dagger}

\begin{document}
\title{Giant field-free transverse Josephson diode effect in  altermagnets}
\author{Bijay Kumar Sahoo}
\affiliation{School of Physics, University of Hyderabad, Prof. C. R. Rao Road, Gachibowli, Hyderabad-500046, India}
\author{ Abhiram Soori }
\email{abhirams@uohyd.ac.in}
\affiliation{School of Physics, University of Hyderabad, Prof. C. R. Rao Road, Gachibowli, Hyderabad-500046, India}

\begin{abstract}
We predict a field-free transverse Josephson diode effect in altermagnets (AMs) with Rashba spin--orbit coupling, achieving diode efficiencies exceeding $3000\%$ and unidirectional transverse supercurrents in four-terminal junctions. In this geometry, a longitudinal phase bias generates transverse supercurrents that exhibit nonreciprocity and a finite anomalous phase shift, while the longitudinal current itself displays a Josephson diode effect. Both responses are tunable via the N\'eel vector orientation. We further show that the effect remains robust against moderate disorder and imperfect interfaces. These results establish AMs as a promising platform for nonreciprocal superconducting transport, with clear routes toward experimental realization.
\end{abstract}

\maketitle

\section{ Introduction}
Nonreciprocal transport, exemplified by the diode effect, is central to next-generation electronics, yet superconducting realizations typically require magnetic fields. A Josephson junction—formed by two superconductors (SCs) separated by a nonsuperconducting region—supports a dissipationless current driven by the phase difference between the SC order parameters. This dependence is captured by the current--phase relation (CPR), whose extrema define the critical currents. When both inversion ($I$) and time-reversal ($TR$) symmetries are broken, the critical currents become directionally asymmetric, giving rise to the Josephson diode effect (JDE), now observed across diverse superconducting platforms~\cite{Ando2020,Baumgartner_2022,turini2022,Kochan2023,Satoshi2023}. JDE also holds promise for energy-efficient computing and current sensing.

 A recently identified class of collinear magnets, altermagnets (AMs), offers a compelling platform for nonreciprocal superconducting transport~\cite{Mazin21,smejkal22c,smejkal22a}. AMs feature spin-polarized Fermi surfaces like ferromagnets but retain vanishing net magnetization as in antiferromagnets. Their opposite-spin sublattices are related by crystal rotations rather than translations, enabling intrinsic breaking of $TR$ symmetry without external fields. These unique properties make AMs particularly suitable for superconducting hybrid devices. Recent studies have predicted unconventional superconducting phenomena in SC/AM junctions~\cite{Chi23,Papaj23,Jabir23,Beenakker23,Wei24,Zhang2024,Lu24,Sun25}, including crystal-orientation-dependent Andreev reflection~\cite{Chi23,Papaj23}, $0$--$\pi$ oscillations without net magnetization~\cite{Jabir23,Zhang2024}, orientation-induced phase shifts~\cite{Beenakker23}, $\phi$-junction behavior~\cite{Lu24}, and non-reciprocal supercurrent transport~\cite{Sayan24}. Furthermore, introducing spin--orbit coupling~(SOC) into AM-based junctions enables anomalous Josephson currents along with JDE tunable by the N\'eel vector~\cite{Wei24}.

\begin{figure}[htb]
    \centering
    \includegraphics[width=8cm]{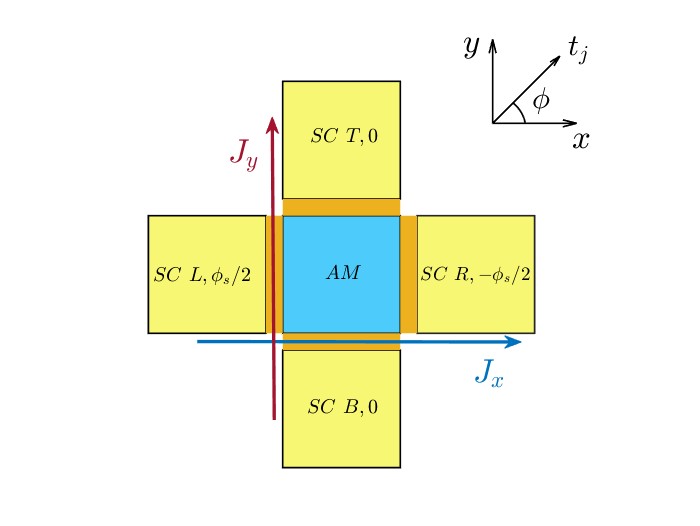}
    \caption{Schematic of the proposed four-terminal junction. The left and right superconducting terminals have phases $\phi_s/2$ and $-\phi_s/2$, respectively, while the top and bottom terminals are held at zero phase. In the central region, the Néel vector of the altermagnet makes an angle $\phi$ with the $x$-axis.}
    \label{fig:schem}
\end{figure}

In a four-terminal Josephson junction, a superconducting phase bias $\phi_s$ applied between the longitudinal leads $L$ and $R$  can generate transverse supercurrents [see Fig.~\ref{fig:schem}], making the transverse current $J_y(\phi_s)$ a well-defined function of the longitudinal phase bias $\phi_s$. We define the transverse critical currents as the maximum and minimum values of $J_y(\phi_s)$, denoted by $J_y^{\max}$ and $J_y^{\min}$. In the absence of a transverse Josephson diode effect (TJDE), the response is reciprocal, with $J_y^{\max}=-J_y^{\min}$, while a deviation from this relation indicates a nonreciprocal transverse response. 
While transverse Josephson currents have been theoretically predicted on topological insulator surfaces under in-plane fields~\cite{Oleksii2021} and TJDE has been proposed in tilted Dirac materials~\cite{Zeng2025}, realistic detection schemes remain lacking. TJDE has also been demonstrated in four-terminal junctions with spin--orbit coupling and an in-plane Zeeman field~\cite{Bijay25}. However, its realization in field-free settings remains largely unexplored.

Here, we show that AMs enable TJDE in a field-free setting when combined with Rashba SOC. We consider a four-terminal junction in which an AM region connects to four $s$-wave superconductors: opposite terminals apply a longitudinal phase bias (with phases $\phi_s/2$ and $-\phi_s/2$), while the transverse terminals are held at zero phase~[Fig.~\ref{fig:schem}]. This phase bias drives finite transverse currents that exhibit JDE, which is tunable by the N\'eel vector, and in certain parameter regimes, these transverse currents become unidirectional. Both longitudinal and transverse diode effects emerge, with the TJDE coefficient exceeding $3000\%$ for realistic parameters. The setup not only realizes the TJDE but also provides a clear route for experimental detection, with the effect persisting up to $T_c$, highlighting feasibility in present multiterminal Josephson platforms~\cite{pankra2020}.

\section{ System}
We consider a four-terminal Josephson junction consisting of a central AM region with Rashba SOC coupled to four superconducting blocks arranged in a cross geometry, as shown in Fig.~\ref{fig:schem}. The system is modeled on a square lattice within tight-binding framework. The total Hamiltonian is
\bea
H &=& H_L + H_{AM} + H_R + H_B + H_T + H_{LA} + H_{RA} \nn \\
&& + H_{TA} + H_{BA},  
\label{eq:H}
\eea
where \(H_{L}, H_{R}, H_{T}, H_{B}\) describe the left (\(L\)), right (\(R\)), top (\(T\)), and bottom (\(B\)) SC blocks, respectively; \(H_{AM}\) describes the central AM region with SOC; and \(H_{pA}\) (\(p=L,R,T,B\)) represent the tunnel couplings between the AM and each SC block. The N\'eel vector of the AM lies in the \(xy\)-plane and makes an angle \(\phi\) with the \(x\)-axis. The left (right) SC blocks are held at superconducting phases \(\phi_s/2\) (\(-\phi_s/2\)), while the top and bottom SC blocks are set to  zero phase. The explicit forms of all Hamiltonian terms are given in Appendix-A.

The Josephson currents \(J_p\) (\(p=L,R,T,B\)) flowing between the AM region and the SC blocks are computed using the Bogoliubov-deGenne formalism described in Appendix-B. Importantly, the system is configured such that if either the SOC or the altermagnetic term \(t_j\) is set to zero, the transverse currents into the top or bottom terminals vanish.

The central AM region with SOC is described in momentum space by
\bea
H_k &=& \ep_k\si_0 + 2t_j(\cos k_xa - \cos k_ya)\si_{\phi} \nn \\
&& + \al(\si_x\sin k_ya - \si_y\sin k_xa),
\label{CAM}
\eea
where $\ep_k=-2t_0(\cos k_xa+\cos k_ya)-\mu_a$, 
$\si_{\phi}=\si_x\cos\phi+\si_y\sin\phi$, and $\si_{x,y,z}$ are Pauli matrices, and $a$ is the lattice constant. 
Here $t_0$ is the hopping strength, $t_j$ the AM strength, and $\alpha$ the SOC strength.  
We model the system as a $20\times 20$ central AM region connected to four $s$-wave SCs of same size as shown in Fig~\ref{fig:schem}.
AMs break TR symmetry intrinsically, mimicking Zeeman fields without net magnetization.  The current flowing from $L$ to $T/B$ due to the phase bias $\phi_s/2$ is equal to the current flowing from $T/B$ to $R$, since the phase difference between $T/B$ and $R$ is also $\phi_s/2$ owing to the symmetry of the setup. Consequently, there is no net current transfer between the longitudinal ($L$, $R$) and transverse ($T$, $B$) terminals, leading to $J_L = J_R \equiv J_x$ and $J_T = J_B \equiv J_y$. 
To quantify diode response, we define the longitudinal and transverse diode coefficients as
\begin{equation}
\gamma_d = \frac{2(J_d^{\max} + J_d^{\min})}{J_d^{\max} - J_d^{\min}}, \quad d = x,y.
\end{equation}

\begin{figure}[htb]
    \includegraphics[width=4.2cm]{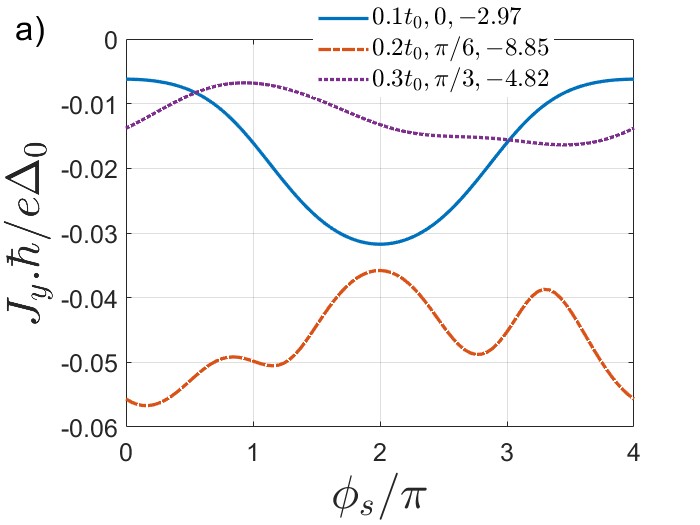}
    \includegraphics[width=4.2cm]{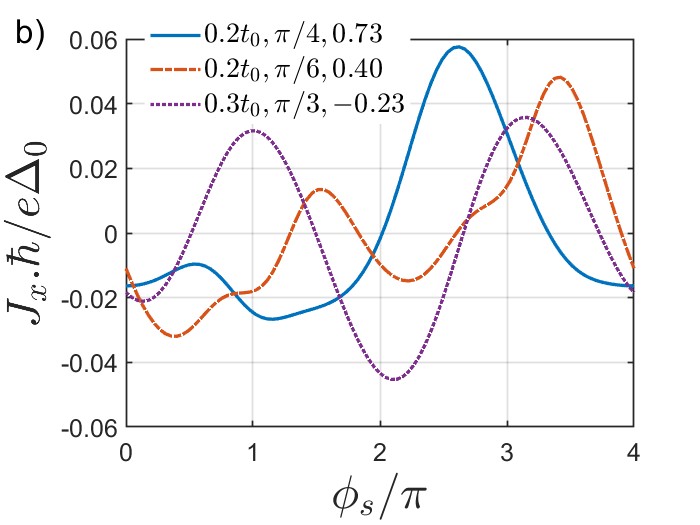}
    \includegraphics[width=4.2cm]{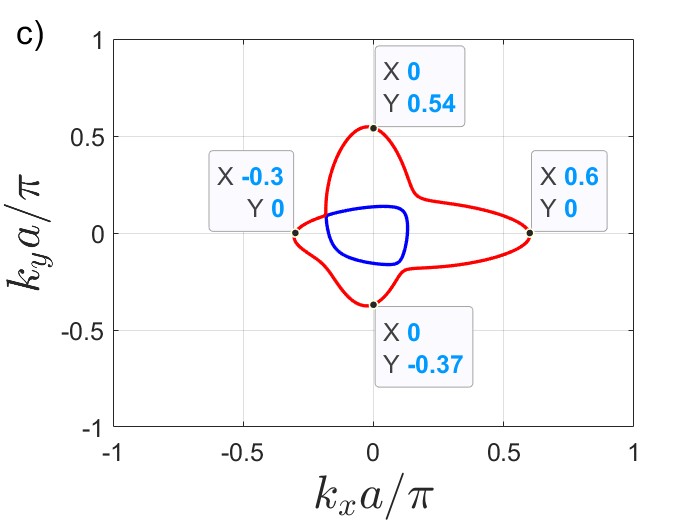}
    \includegraphics[width=4.2cm]{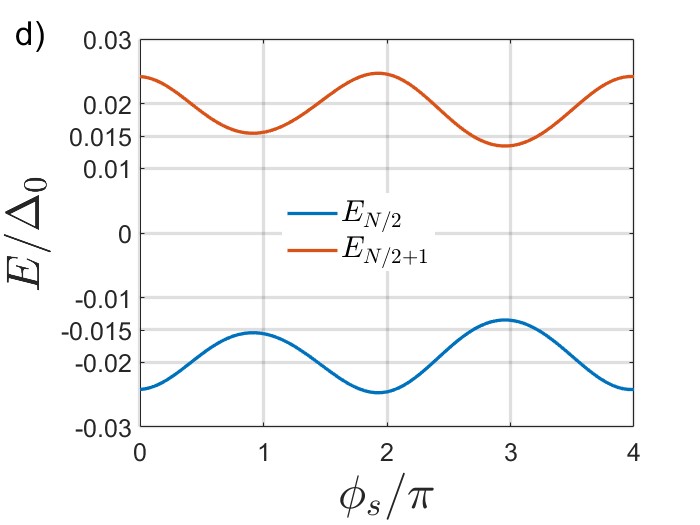}
    \caption{CPR: Transverse Josephson current ($J_y$) in (a) and Longitudinal Josephson current ($J_x$) in (b) for different values of $\al$ and $\phi$ with respective diode coefficients as indicated in the legend by $(\al,\phi,\ga)$, (c) Fermi surface of AM, (d) Energies of states $j=N/2,N/2+1$ ($N\times N$ is the size of the matrix $H$, where $N=4[4L_s^xL_s^y+L_a^xL_a^y]$) versus superconducting phase difference $\phi_s$. Parameters: $\mu_s=\mu_a=-3.6t_0$, $\De_0=0.06t_0$, $t_j=0.75t_0$ for all figures, and $\al=0.3t_0$, $\phi=\pi/3$ for Fig. (c) and (d).The number of sites in $x$ and $y$-directions for SC ($L_s^x,~L_s^y$) and AM ($L_a^x,~L_a^y$) regions are  $L_s^x=L_s^y=L_a^x=L_a^y=20$.}
    \label{fig:CPR}
\end{figure}

To connect our model to realistic materials, we adopt parameter values motivated by density functional theory   studies of the altermagnetic candidate KRu$_4$O$_8$~\cite{Weber24}. In particular, the relative strengths of the altermagnetic exchange and Rashba spin--orbit coupling are chosen to be consistent with its bandstructure, ensuring that the parameter regimes exhibiting unidirectional transverse currents are physically realistic.
Specifically, we consider parameters relevant to KRu$_4$O$_8$~\cite{Weber24,Knolle25}: $t_0=51$\,meV, $t_j=0.75t_0$, $t'=0.5t_0$, and $\mu_s=\mu_a=-3.6t_0$, with Rashba coupling $\alpha \sim 0.1t_0$--$0.3t_0$~\cite{Diniz24,Li25,Chen2025}. The superconducting pairing strength is taken as $\Delta_0=0.06t_0$, corresponding to NbN~\cite{Duc1987}.
\section{Results}
\subsection{ Current-phase relations} Figures~\ref{fig:CPR}(a,b) show the CPRs of $J_y$ and $J_x$ for various $\alpha$ and $\phi$. 
The CPRs display a $4\pi$-periodicity because three distinct SC phases ($\phi_s/2, 0, -\phi_s/2$) are involved; this is confirmed by the $4\pi$-periodic spectrum of the energy levels $E_{N/2}$ and $E_{N/2+1}$ [Fig.~\ref{fig:CPR}(d)].  Though the phase difference between the left and the right SC's is $\phi_s$, the phase difference between the left and the top/bottom SCs is $\phi_s/2$, making the CPR $4\pi$ periodic in $\phi_s$. 
Both transverse and longitudinal CPRs exhibit anomalous Josephson effect (AJE) and JDE. 
Breaking of $k_y\!\leftrightarrow\!-k_y$ symmetry for $\phi\neq\pi/2,3\pi/2$ induces AJE and JDE in $J_y$, while breaking of $k_x\!\leftrightarrow\!-k_x$ symmetry for $\phi\neq0,\pi$ gives JDE and AJE in $J_x$.  
The resulting Fermi surface asymmetry for $\phi=\pi/3$ is shown in Fig.~\ref{fig:CPR}(c).  
For $\alpha=0.2t_0$ and $\phi=\pi/4$, the longitudinal diode coefficient reaches $\sim73\%$, and for some parameter sets $J_y$ becomes strictly unidirectional [Fig.~\ref{fig:CPR}(a)].
One way to understand the AJE and JDE is to note that, in a single back-and-forth traversal, the total phases accumulated by two sets of processes—(i) an electron moving right and a hole moving left, and (ii) an electron moving left and a hole moving right—differ in magnitude due to the broken symmetry between $k_x$ and $-k_x$. A similar reasoning applies to the transverse direction, accounting for the AJE and TJDE.

 \begin{figure}[htb]
    \centering
    \includegraphics[width=4.2cm]{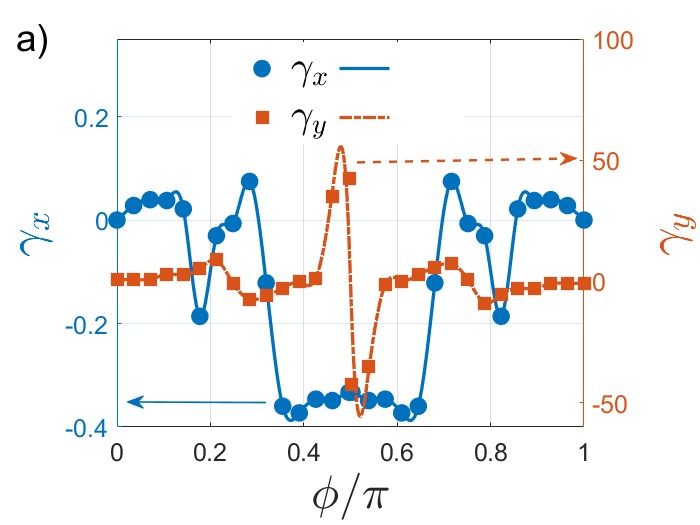}
    \includegraphics[width=4.2cm]{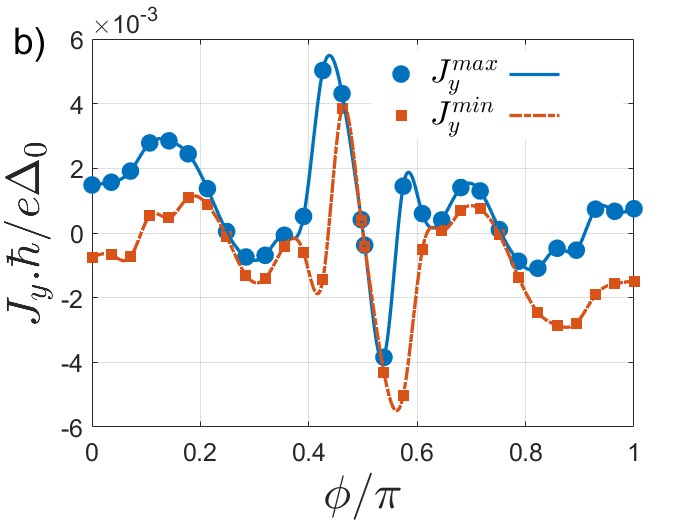}
    \caption{ Transverse (Longitudinal) diode effect coefficient $\ga_y$ ($\ga_x$) versus $\phi$ in (a) and transverse Josephson critical currents versus $\phi$ in (b) for $\al=0.3t_0$ and all other parameters are the same as in Fig.~\ref{fig:CPR}.}
    \label{fig:dcvsphi}
\end{figure}

\subsection{ Diode effect coefficients versus $\phi$} 
In Fig.~\ref{fig:dcvsphi}(a) and (b), we plot $\gamma_{x,y}$ and the transverse critical currents ($J_y^{\max}, J_y^{\min}$) as functions of $\phi$. The solid lines correspond to spline fits serving as a guide to the eye, while the filled circles and squares denote the actual data points. Here we have shown the range of $\phi=0$ to $\pi$, because as $\phi \to \pi+\phi$, $\ga_{x,y} \to -\ga_{x,y}$. This can be understood from the Hamiltonian in equation \ref{CAM}. Under $\phi \to \pi+\phi$, $\si_{\phi} \to {-\si_{\phi}}$ which means the direction of N\'eel vector is reversed in the plane. This leads to reversal in sign of $\ga_{x,y}$, under $\phi \to\pi+\phi$.  As expected, $\gamma_y$ vanishes at $\phi=\pi/2,3\pi/2$ (due to restored $k_y\to -k_y$ symmetry) and $\gamma_x$ at $\phi=0,\pi$ (due to restored $k_x \to -k_x$ symmetry). In Fig.~\ref{fig:dcvsphi} (a), we can see that the $\ga_y$ is of the order of $100\%$ and for some values of $\phi$ it exceeds $3000\%$, while $\ga_x$ close to $40\%$ for $\al=0.3t_0$.
This significantly surpasses the diode effect coefficients of roughly $30\%$ reported in SOC systems with a Zeeman field~\cite{Kochan2023}.
Figure~\ref{fig:dcvsphi}(b) shows the corresponding critical currents $J_y^{\max,\min}$, which also vanish at $\phi=\pi/2,3\pi/2$ and exhibit unidirectionality for some ranges (e.g. for $\phi\in[0.1\pi,0.35\pi]$ and $[0.65\pi,0.9\pi]$).

\begin{figure}[htb]
    \centering
    \includegraphics[width=4.2cm]{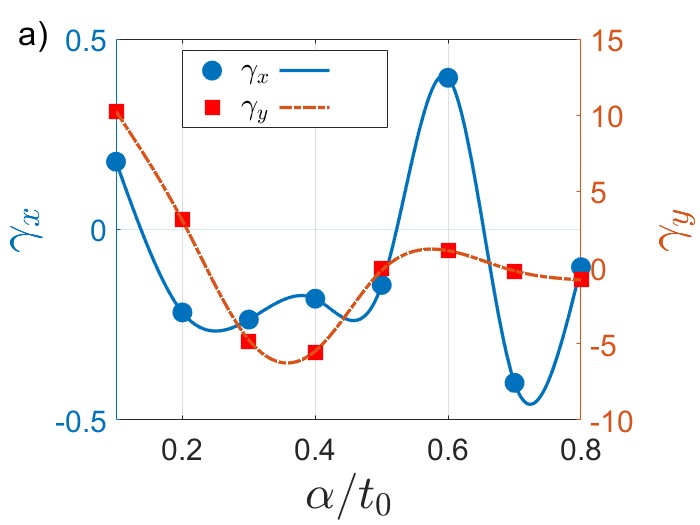}
    \includegraphics[width=4.2cm]{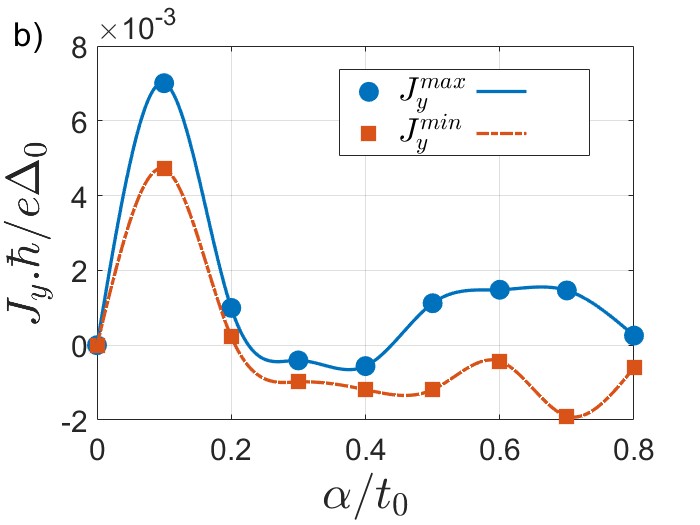}
    \caption{(a) Transverse (Longitudinal) diode effect coefficient $\gamma_y$ ($\gamma_x$) versus SOC strength $\al$, (b) Transverse Josephson critical currents ($J_y^{max}$ and $J_y^{min}$) versus $\al$,  for $\phi=\pi/3$ and other parameters are the same as in Fig.~\ref{fig:CPR}}
    \label{fig:dcvsal}
\end{figure}

\subsection{ Diode effect coefficients versus SOC strength} The strength of SOC is gate tunable~\cite{ast2008}.
Figure~\ref{fig:dcvsal}(a,b) shows $\gamma_{x,y}$ and $J_y^{\max,\min}$ versus $\alpha$ at $\phi=\pi/3$.  
At $\alpha=0$, both $\gamma_x$ and $J_y$ vanish due to restored momentum symmetry. For $\alpha \in [0,0.4t_0]$, $J_y$ becomes unidirectional.

\subsection{ Current--phase relations with disorder}
\begin{figure}
    \centering
    \includegraphics[width=4.2cm]{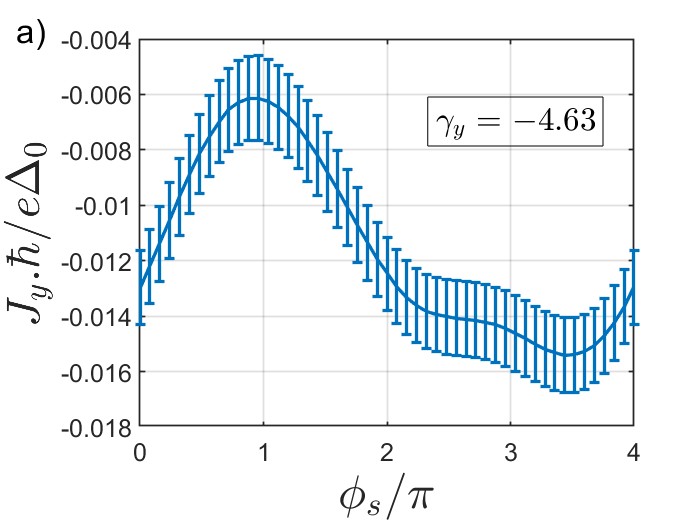}
    \includegraphics[width=4.2cm]{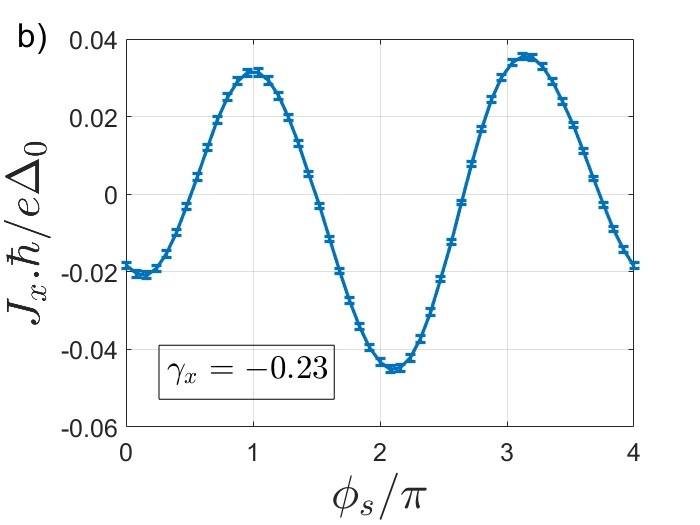}
    \caption{CPR with bond and onsite disorder: Average transverse Josephson current ($J_y$) in (a) and Average longitudinal Josephson current ($J_x$) in (b) for $10$ disorder relaizations. The errorbars show standard deviation.  Parameters: $\alpha=0.3t_0,~\phi=\pi/3,~w_b=w_0=0.2\De_0$ and other parameters are same as in Fig.~\ref{fig:CPR}.}
    \label{fig:CPRd}
\end{figure}

We examine the current--phase relations (CPRs) of both transverse and longitudinal Josephson currents in the presence of disorder, as shown in Fig.~\ref{fig:CPRd}, for $\alpha=0.3t_0$, $\phi=\pi/3$, and $L_{s/a}^d=20$ (with $d=x,y$). We include both on-site disorder in the AM region and bond disorder at the interfaces connecting the AM to the SC leads. The on-site disorder is chosen randomly from the range $[-w_0,w_0]$, while the hopping amplitudes at the AM--SC interfaces are varied randomly within $[t'-w_b,\,t'+w_b]$. Here, we take $w_0=w_b=0.2\Delta_0$.
The Josephson current is averaged over $10$ disorder realizations, with error bars indicating the standard deviation. As shown in Fig.~\ref{fig:CPRd}(a), the unidirectional nature of the transverse Josephson current remains robust in the presence of disorder, with $\gamma_y$ exceeding $400\%$ for the chosen parameters. Although the overall magnitude of the current is reduced by approximately an order of magnitude compared to the clean case [Fig.~\ref{fig:CPR}(a)], the nonreciprocal character persists. Similarly, the longitudinal CPR continues to exhibit a diode effect, with the efficiency $\gamma_x$ exceeding $20\%$ for the same parameter set.

\begin{figure}[htb]
    \centering
    \includegraphics[width=4.2cm]{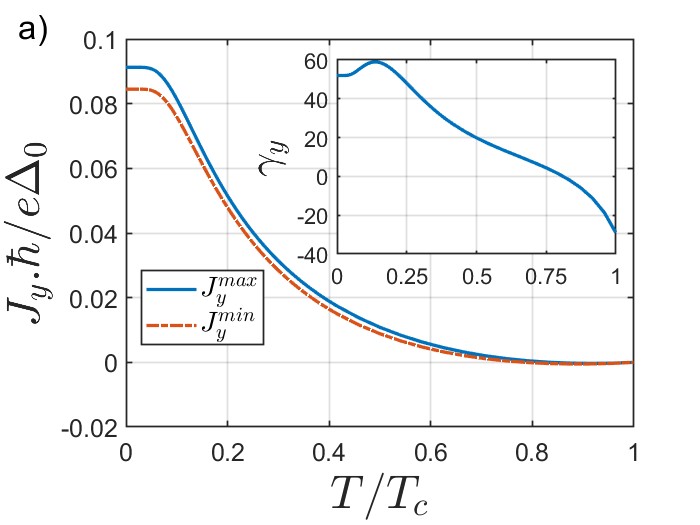}
    \includegraphics[width=4.2cm]{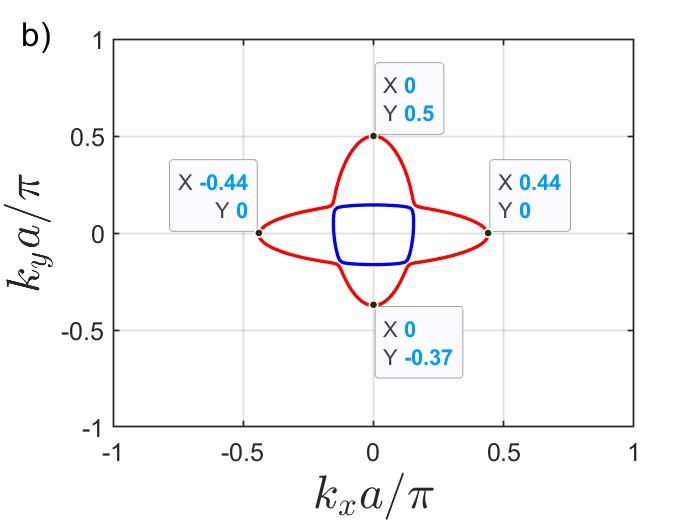}
    \includegraphics[width=4.2cm]{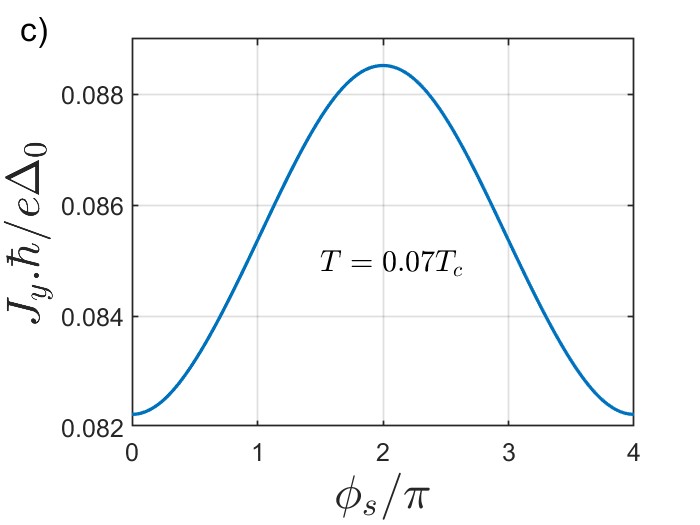}
    \includegraphics[width=4.2cm]{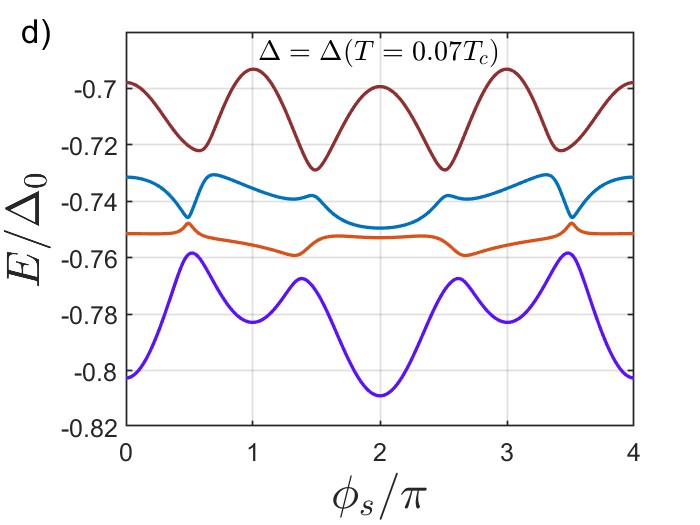}
    \caption{(a) Transverse critical currents versus temperature $T$, inset: transverse diode effect coefficient $\ga_y$ versus $T$ (b)~Fermi surface of AM (c) CPR for $J_y$ (d) Andreev bound state energies versus the phase bias. Parameters: $\al=0.1t_0$, $\phi=0$, $L_s^d=L_a^d=10$ with $d=x,y$ and other parameters are the same as in Fig.~\ref{fig:CPR}.}
    \label{fig:dcvsTemp}
\end{figure} 

\subsection{ Finite temperatures}
Taking $T_c=12\,{\rm K}$ for NbN~\cite{Leith2021}, we compute the finite-temperature Josephson current as 
$J_p(T)=\sum_j \bra{u_j}\hat J_p\ket{u_j}\tanh(E_j/2k_BT)$, 
where the sum runs over negative-energy states, $\ket{u_j}$ is the $j$-th eigenstate, and $\hat J_p$ is the current operator at the AM/$p$-SC interface ($p=L,R,T,B$). The superconducting gap depends on $T$ via $\De(T)=\De_0\tanh[1.74\sqrt{T_c/T-1}]$~\cite{Jabir23}.  

Figure~\ref{fig:dcvsTemp}(a) shows the transverse critical currents as functions of temperature $T$, with the inset displaying the diode coefficient $\gamma_y(T)$ for $\alpha=0.1t_0$, $\phi=0$, and $L_{s/a}^{x,y}=10$. The unidirectional behavior originates from the strong $k_y$ asymmetry of the AM Fermi surface [Fig.~\ref{fig:dcvsTemp}(b)]. As $T$ increases, the critical currents decrease and vanish as $T\to T_c$, reflecting the closing of the superconducting gap. In contrast, $\gamma_y$ exhibits a nonmonotonic dependence on $T$.

For $T\ll T_c$, $J_y^{\rm max}$, $J_y^{\rm min}$, and $\gamma_y$ remain nearly constant since thermal excitation of quasiparticles is negligible. When $k_BT\sim\Delta_0$ ($T\approx0.07T_c$), $J_y^{\rm max}$ and $J_y^{\rm min}$ evolve at different rates, leading to a peak in $\gamma_y(T)$. This behavior originates from the unequal spacing of quasiparticle levels near the phase values where $J_y$ is extremal [Figs.~\ref{fig:dcvsTemp}(c,d)]: near $\phi_s=0$, $J_y\approx J_y^{\rm min}$, while near $\phi_s=2\pi$, $J_y\approx J_y^{\rm max}$. Figure~\ref{fig:dcvsTemp}(d) shows a subset of negative-energy Andreev levels close to zero energy; by particle--hole symmetry, each level $-E_j$ is accompanied by $E_j$. The level spacings differ between $\phi_s=0$ and $\phi_s=2\pi$, causing $J_y^{\rm min}$ to be suppressed at lower temperatures than $J_y^{\rm max}$. As $T\to T_c$, $\gamma_y$ approaches a finite value ($\approx -29$); however, since the absolute currents vanish in this limit, the diode effect loses practical significance.

\section{ Discussion and Conclusion }
We demonstrate that a four-terminal Josephson architecture with a central altermagnetic region featuring Rashba SOC, connected to four $s$-wave superconducting contacts, exhibits rich transverse responses, including unidirectional currents, alongside longitudinal AJE and JDE.
Applying a phase difference across the longitudinal terminals was found to induce supercurrents not only along the direction of phase bias but also in the transverse direction. These transverse currents show two striking features: they persist even at zero phase offset (signaling an anomalous Josephson effect), and their critical values are unequal in magnitude, evidencing a diode-like behavior. Such nonreciprocal characteristics emerge from lifting of degeneracies between opposite-momentum states: imbalance between $k_x$ and $-k_x$ modes gives rise to longitudinal nonreciprocity, while imbalance between $k_y$ and $-k_y$ modes generates transverse currents with both anomalous and diode responses. In fact, the calculated transverse diode efficiency reaches values exceeding $3000\%$, and for certain parameter regimes, the current–phase relation becomes strongly unidirectional. Giant TJDE can be utilized for current sensing. Interestingly, both JDE and AJE are present even in the presence of both and on-site disorder (for the strength of disorder small compared to $\Delta_0$). We observe $\ga_y$ exceeding $400\%$, while $\ga_x$ reaches more than $20\%$ for disorder strength $w_0=w_b=0.2\Delta_0$. More importantly, the transverse Josephson current remains unidirectional even in the presence of disorder.

We have also performed calculations for conventional collinear antiferromagnets (AFMs) with Rashba SOC and do not observe TJDE. The key distinction between AFMs and  AMs lies in their band structure: while AFMs remain spin-degenerate in the absence of additional symmetry breaking, AMs exhibit intrinsic, momentum-dependent spin splitting even without SOC. 
The emergence of TJDE requires not only the breaking of time-reversal ($TR$) and inversion ($I$) symmetries, but also the breaking of opposite-momentum symmetry, i.e., $E(k_y)\neq E(-k_y)$. In our system, $TR$ symmetry is broken by the altermagnetic order, $I$ symmetry by Rashba SOC, and the in-plane orientation of the N\'eel vector further lowers the symmetry to generate a transverse momentum asymmetry. This combination leads to the observed nonreciprocal transverse response. 
In contrast, conventional AFMs with Rashba SOC typically retain additional symmetries (such as combined $PT$ symmetry) that enforce $E(k_y)=E(-k_y)$, thereby suppressing the TJDE. Thus, AMs provide a natural and field-free platform for realizing transverse nonreciprocal superconducting transport.

At finite temperatures, the critical currents decrease and vanish as $T \!\to\! T_c$, yet the transverse currents retain their unidirectional character. The diode effect coefficient remains finite throughout the entire temperature range from $0$ to $T_c$.

The proposed four-terminal Josephson geometry provides a direct route for experimental detection. A phase bias applied between the longitudinal leads (left and right) generates transverse supercurrents in the top and bottom leads, which can be measured using standard superconducting transport techniques. The unidirectional nature of the transverse current can be identified by changing the longitudinal phase bias and monitoring the corresponding change in the transverse response. 

A key experimental challenge in realizing this setup is the control of the N\'eel vector in AMs, which governs the magnitude and direction of the effect. While several approaches have been demonstrated for manipulating the N\'eel vector in antiferromagnets~\cite{zhang2022,Godinho2018,Mahmood2021}, only a few recent works have explored analogous control schemes in AMs~\cite{han2025,Amit26}. In particular, Ref.~\cite{Amit26} proposes a deterministic method to switch the N\'eel vector in AMs. The development of reliable control protocols will therefore be crucial for tuning and optimizing the transverse Josephson diode effect.

The four-terminal cross geometry considered here represents the minimal setup to define and probe transverse Josephson currents independently of the longitudinal ones. The transverse Josephson diode effect (TJDE) identified in this work is not specific to this geometry, but arises from the momentum-space asymmetry of the quasiparticle spectrum induced by the interplay of altermagnetism, Rashba spin--orbit coupling, and the orientation of the N\'eel vector. Consequently, the transverse diode response is expected to persist in more general multiterminal geometries ($N>4$), provided transverse current paths are available. While additional terminals may modify the magnitude of the effect through changes in current distribution and interference between transport channels, the underlying mechanism remains unchanged, indicating that the TJDE is not a geometric interference effect tied to the cross geometry.

Recent advances in multiterminal Josephson junctions~\cite{pankra2020,Gupta2023} provide a promising platform to probe unconventional superconducting responses arising from engineered symmetry breaking~\cite{frolov2004,glick2018}. In this context, Ref.~\cite{Sun26} proposes a current-driven superconducting diode effect in a multiterminal disk-like geometry, where unidirectional transverse currents emerge when the longitudinal current exceeds a critical value. These developments highlight the experimental feasibility of nonreciprocal superconducting transport in multiterminal devices.

A crucial advantage of this platform is that it requires no external magnetic field: the combination of SOC and the N\'eel vector of the altermagnet naturally breaks the symmetries needed to produce nonreciprocal superconducting transport, while preserving zero net magnetization. The phenomena discussed in this work extend to other AMs and SOC types enabling field-free diodes in quantum circuits. This work therefore establishes altermagnets as a fertile setting for realizing the transverse Josephson diode effect and transverse anomalous Josephson effect, opening a new route toward field-free nonreciprocal superconducting electronics and bridging altermagnetism with device physics.

\begin{acknowledgments}
We thank Amit Agarwal and Udit Khanna for comments on the manuscript. Computational facility was provided from the University of Hyderabad Institute of Eminence Grant No.UH/RITE/PHY/SS/IoE-RC522020/01. We thank Soma Sanyal for computational support. AS thanks Science and Engineering Research Board (now Anusandhan National Research Foundation) Core Research grant (CRG/2022/004311) for financial support. BKS thanks the Ministry of Social Justice and Empowerment, Government of India, for a fellowship through NFOBC.
\end{acknowledgments}
\bibliography{ref_almag}

\begin{widetext}

\section*{ Appendix A: Hamiltonian  }
Different contributions to the Hamiltonian in eq.~\eqref{eq:H} are summarized below.
\bea   
H_L &=& \sum_{n_x=1}^{L_s^x-1} \sum_{n_y=L_s^y+1}^{L_{sa}^y} \Big[ -t_0 (\Psi_{n_x+1,n_y}^{\dg}\tau_z \Psi_{n_x,n_y}+h.c.)\Big] +\sum_{n_x=1}^{L_s^x} \sum_{n_y=L_s^y+1}^{L_{sa}^y-1}\Big[ -t_0 (\Psi_{n_x,n_y+1}^{\dg}\tau_z \Psi_{n_x,n_y}+h.c.) \Big]\nn \\
&& -\mu_s \sum_{n_x=1}^{L_s^x} \sum_{n_y=L_s^y+1}^{L_{sa}^y}\Psi_{n_x,n_y}^{\dg}\tau_z \Psi_{n_x,n_y} -\De_0 \sum_{n_x=1}^{L_s^x} \sum_{n_y=L_s^y+1}^{L_{sa}^y}\Psi_{n_x,n_y}^{\dg} (\cos \phi_l ~\tau_y\si_y+\sin \phi_l ~\tau_x\si_y) \Psi_{n_x,n_y}, \nn \\
H_{AM} &=& \sum_{n_x=L_s^x+1}^{L_{sa}^x-1} \sum_{n_y=L_s^y+1}^{L_{sa}^y} \Big[ -t_0 (\Psi_{n_x+1,n_y}^{\dg}\tau_z \Psi_{n_x,n_y}+h.c.)\Big] +\sum_{n_x=L_s^x+1}^{L_{sa}^x} \sum_{n_y=L_s^y+1}^{L_{sa}^y-1}\Big[ -t_0 (\Psi_{n_x,n_y+1}^{\dg}\tau_z \Psi_{n_x,n_y}+h.c.) \Big]\nn \\
&& -\mu_a \sum_{n_x=L_s^x+1}^{L_{sa}^x}\sum_{n_y=L_s^y+1}^{L_{sa}^y}\Psi_{n_x,n_y}^{\dg}\tau_z \Psi_{n_x,n_y}+ t_j \sum_{n_x=L_s^x+1}^{L_{sa}^x-1} \sum_{n_y=L_s^y+1}^{L_{sa}^y}\Big[\Psi_{n_{x+1},n_y}^{\dg}(\cos \phi ~\tau_z\si_x+\sin \phi ~\tau_0\si_y)\Psi_{n_x,n_y}\Big]\nn \\
&& \nn \\
&& -t_j\sum_{n_x=L_s^x+1}^{L_{sa}^x} \sum_{n_y=L_s^y+1}^{L_{sa}^y-1}\Big[\Psi_{n_{x},n_{y+1}}^{\dg}(\cos \phi ~\tau_z\si_x+\sin \phi ~\tau_0\si_y)\Psi_{n_x,n_y}\Big]+ \frac{\al}{2} \sum_{n_x=L_s^x+1}^{L_{sa}^x}\sum_{n_y=L_s^y+1}^{L_{sa}^y-1}(i\Psi_{n_x,n_y+1}^{\dg}\tau_0\si_x \Psi_{n_x,n_y}+h.c.)\nn \\
&&-\frac{\al}{2}\sum_{n_x=L_s^x+1}^{L_{sa}^x-1} \sum_{n_y=L_s^y+1}^{L_{sa}^y}(i\Psi_{n_x+1,n_y}^{\dg}\tau_z\si_y \Psi_{n_x,n_y}+h.c.),   \nn \\
H_R &=& \sum_{n_x=L_{sa}^x+1}^{L_{sas}^x-1} \sum_{n_y=L_s^y+1}^{L_{sa}^y} \Big[ -t_0 (\Psi_{n_x+1,n_y}^{\dg}\tau_z \Psi_{n_x,n_y}+h.c.)\Big] +\sum_{n_x=L_{sa}^x+1}^{L_{sas}^x} \sum_{n_y=L_s^y+1}^{L_{sa}^y-1}\Big[ -t_0 (\Psi_{n_x,n_y+1}^{\dg}\tau_z \Psi_{n_x,n_y}+h.c.) \Big] \nn \\
&& -\mu_s \sum_{n_x=L_{sa}^x+1}^{L_{sas}^x}\sum_{n_y=L_s^y+1}^{L_{sa}^y}\Psi_{n_x,n_y}^{\dg}\tau_z \Psi_{n_x,n_y} -\De_0 \sum_{n_x=L_{sa}^x+1}^{L_{sas}^x}\sum_{n_y=L_s^y+1}^{L_{sa}^y}\Psi_{n_x,n_y}^{\dg} (\cos \phi_r ~\tau_y\si_y+\sin \phi_r ~\tau_x\si_y) \Psi_{n_x,n_y}, \nn \\
H_B &=& \sum_{n_x=L_{s}^x+1}^{L_{sa}^x-1} \sum_{n_y=1}^{L_{s}^y}\Big[ -t_0 (\Psi_{n_x+1,n_y}^{\dg}\tau_z \Psi_{n_x,n_y}+h.c.)\Big]+\sum_{n_x=L_{s}^x}^{L_{sa}^x} \sum_{n_y=1}^{L_{s}^y-1}\Big[ -t_0 (\Psi_{n_x,n_y+1}^{\dg}\tau_z \Psi_{n_x,n_y}+h.c.) \Big]\nn \\ && -\mu_s \sum_{n_x=L_{s}^x+1}^{L_{sa}^x}\sum_{n_y=1}^{L_{s}^y}\Psi_{n_x,n_y}^{\dg}\tau_z \Psi_{n_x,n_y}-\De_0 \sum_{n_x=L_{s}^x+1}^{L_{sa}^x}\sum_{n_y=1}^{L_{s}^y}\Psi_{n_x,n_y}^{\dg} \tau_y \si_y \Psi_{n_x,n_y},\nn \\
H_T &=& \sum_{n_x=L_{s}^x+1}^{L_{sa}^x-1} \sum_{n_y=L_{sa}^y+1}^{L_{sas}^y} \Big[ -t_0 (\Psi_{n_x+1,n_y}^{\dg}\tau_z \Psi_{n_x,n_y}+h.c.)\Big]+\sum_{n_x=L_{s}^x}^{L_{sa}^x} \sum_{n_y=L_{sa}^y+1}^{L_{sas}^y-1}\Big[ -t_0 (\Psi_{n_x,n_y+1}^{\dg}\tau_z \Psi_{n_x,n_y}+h.c.) \Big]\nn \\ && -\mu_s \sum_{n_x=L_{s}^x+1}^{L_{sa}^x}\sum_{n_y=L_{sa}^y+1}^{L_{sas}^y}\Psi_{n_x,n_y}^{\dg}\tau_z \Psi_{n_x,n_y}-\De_0 \sum_{n_x=L_{s}^x+1}^{L_{sa}^x}\sum_{n_y=L_{sa}^y+1}^{L_{sas}^y}\Psi_{n_x,n_y}^{\dg} \tau_y \si_y \Psi_{n_x,n_y}, \nn \eea 

\end{widetext}

\bea
H_{LA} &=& -t' \sum_{n_y=L_s^y+1}^{L_{sa}^y}(\Psi_{L_s+1,n_y}^{\dg} \tau_z \Psi_{L_s,n_y}+h.c.), \nn \\
H_{RA} &=& -t' \sum_{n_y=L_s^y+1}^{L_{sa}^y}(\Psi_{L_{sa}+1,n_y}^{\dg} \tau_z \Psi_{L_{sa},n_y}+h.c.), \nn \\
H_{TA} &=& -t'\sum_{n_x=L_{s}^x+1}^{L_{sa}^x} (\Psi_{n_x,L_{sa}+1}^{\dg} \tau_z \Psi_{n_x,L_{sa}}+h.c.), \nn \\
H_{BA} &=& -t'\sum_{n_x=L_{s}^x+1}^{L_{sa}^x} (\Psi_{n_x,L_{s}+1}^{\dg} \tau_z \Psi_{n_x,L_{s}}+h.c.), 
\eea
Here, $L_s^d$ ($L_a^d$) denotes the number of sites in the SC (AM) regions along the direction $d$ (with $d = x, y$). And $L_{sa}^d = L_s^d + L_a^d$ and $L_{sas}^d = 2L_s^d + L_a^d$.  
\begin{equation}
\Psi_{n_x,n_y} = \begin{bmatrix} c_{n_x,n_y,\uparrow} & c_{n_x,n_y,\downarrow} & c_{n_x,n_y,\uparrow}^\dagger & c_{n_x,n_y,\downarrow}^\dagger \end{bmatrix}^T,
\end{equation}  
where $c_{n_x,n_y,\sigma}$ annihilates an electron with spin $\sigma$ at site $(n_x,n_y)$. The Pauli matrices $\tau_{x,y,z}$ and $\sigma_{x,y,z}$ act in particle-hole and spin spaces, respectively. Other model parameters are as follows: (i)~$t_0$: hopping amplitude within SC and AM regions,
(ii)~$t'$: hopping connecting SC and AM regions, set to $t_0/2$ in our calculations,
(iii)~$t_j$: amplitude of the altermagnetic term,
(iv)~$\De_0$: superconducting pairing potential,
(v)~$\phi_l$ and $\phi_r$ are the phases of the two SCs set to $\phi_s/2$ and $-\phi_s/2$ respectively, 
(vi)~$\alpha$: spin-orbit coupling strength,
(vii)~$\phi$: angle made by N\'eel vector of AM with respect to the $x$-axis,
(viii)~$\mu_s$ ($\mu_a$): chemical potential in SC (AM) regions.

\section*{ Appendix B: Josephson currents }
The equilibrium Josephson current is obtained by summing contributions from all occupied states~\cite{Bijay25}. Charge conservation in the SOC region allows us to define the current operators at the four interfaces connecting the SC and AM regions:
\begin{align}  
\hat{J_L} &= \frac{iet'}{\hbar} \sum_{n_y=L_{s}^y+1}^{L_{sa}^y} (\Psi_{L_s+1,n_y}^\dagger \Psi_{L_s,n_y} - \text{h.c.}) \\
\hat{J_R} &= \frac{iet'}{\hbar} \sum_{n_y=L_{s}^y+1}^{L_{sa}^y} (\Psi_{L_{sa}+1,n_y}^\dagger \Psi_{L_{sa},n_y} - \text{h.c.}) \\
\hat{J_B} &= \frac{iet'}{\hbar} \sum_{n_x=L_{s}^x+1}^{L_{sa}^x} (\Psi_{n_x,L_s+1}^\dagger \Psi_{n_x,L_s} - \text{h.c.}) \\
\hat{J_T} &= \frac{iet'}{\hbar} \sum_{n_x=L_{s}^x+1}^{L_{sa}^x} (\Psi_{n_x,L_{sa}+1}^\dagger \Psi_{n_x,L_{sa}} - \text{h.c.})  
\end{align}  

For each value of $\phi_s$, we numerically diagonalize the Hamiltonian to obtain eigenstates and eigenenergies $(|u_j\rangle, E_j)$. All negative-energy states are assumed filled and positive-energy states empty. 

The total Josephson current is given by  
$ J_p = \sum_j \langle u_j | \hat{J_p} | u_j \rangle,  $
where $p = L, R, T, B$ and $j$ is summed over occupied states.

\section*{Appendix C: Disordered System }
The Hamiltonian describing the system with disorder is as follows
\bea
H &=& H_L + H_{AM} + H_R + H_B + H_T + H_{LA}^D+ H_{RA}^D \nn \\
&& + H_{TA}^D + H_{BA}^D+H_D,
\eea
where the first five terms of the Hamiltonian have the same form as in the clean case, while the other terms are as follows
\bea
H_D &=& \sum_{n_x=L_s^x+1}^{L_{sa}^x}\sum_{n_y=L_s^y+1}^{L_{sa}^y} \epsilon_{n_x,n_y}\Psi_{n_x,n_y}^{\dg}\tau_z \Psi_{n_x,n_y}, \nn \\
 H_{LA}^D &=& \sum_{n_y=L_s^y+1}^{L_{sa}^y}(-t'+w_{L,n_y})(\Psi_{L_s+1,n_y}^{\dg}\tau_z \Psi_{L_s,n_y}+h.c.), \nn \\
H_{RA}^D &=&  \sum_{n_y=L_s^y+1}^{L_{sa}^y} (-t'+w_{R,n_y})(\Psi_{L_{sa}+1,n_y}^{\dg} \tau_z \Psi_{L_{sa},n_y}+h.c.), \nn \\
H_{TA}^D &=& \sum_{n_x=L_{s}^x+1}^{L_{sa}^x} (-t'+w_{T,n_x})(\Psi_{n_x,L_{sa}+1}^{\dg} \tau_z \Psi_{n_x,L_{sa}}+h.c.), \nn \\
H_{BA}^D &=& \sum_{n_x=L_{s}^x+1}^{L_{sa}^x} (-t'+w_{B,n_x}) (\Psi_{n_x,L_{s}+1}^{\dg} \tau_z \Psi_{n_x,L_{s}}+h.c.),~~~~~~~~~
\eea 
 with $\epsilon_{n_x,n_y}$ choosen randomly between $[-w_0,w_0]$, while $w_{L/R/T/B,n_x/n_y}$ choosen randomly between $[-w_b,w_b]$. Here $w_0$ and $w_b$ are the on-site and bond disorder strength respectively.

We employ an approximation in the numerical diagonalization of the Hamiltonian to obtain the results shown in Figs.~\ref{fig:dcvsphi},~\ref{fig:dcvsal}, and~\ref{fig:CPRd}. In the exact treatment, all negative-energy states are assumed to be occupied up to zero energy. However, the Josephson current is primarily determined by states in the vicinity of zero energy, while contributions from states far below zero energy are negligible. Accordingly, we restrict the diagonalization to a subset of eigenstates near zero energy, selecting a number of states equal to twice the number of sites in the central AM region. These states capture the dominant contribution to the current. We have verified the validity of this approximation by reproducing the CPR obtained from exact diagonalization for the same system size.

\end{document}